\documentclass[twocolumn,showpacs,preprintnumbers,amsmath,amssymb]{revtex4}

\usepackage{graphicx,subfigure,epsfig}
\usepackage{dcolumn,amsmath}
\usepackage{bm}

\begin{document}

\preprint{APS/123-QED}

\title{Squeezed States and Affleck Dine Baryogenesis}% Force line breaks with \\

\author{K. V. S. Shiv Chaitanya }
 \email{ph03ph13@uohyd.ernet.in}
 
\author{Bindu A Bambah}%
 \email{bbsp@uohyd.ernet.in}
\affiliation{%
School of Physics, University of Hyderabad, Hyderabad, A.P. 500 046, India.\\
}
\date{\today}
\begin{abstract}
Quantum fluctuations in the post inflationary
Affleck-Dine baryogenesis model are studied.
The squeezed states formalism is used to
give evolution equations for the particle and
anti-particle modes in the early universe. The role of expansion and 
parametric amplification of the  quantum fluctuations on the
baryon asymmetry produced is investigated.
\end{abstract}

\maketitle

\section{\label{sec:level1}Introduction}

The dominance of matter over antimatter is known as baryon
asymmetry. The generation of baryon asymmetry from an initially
symmetric universe is an open problem. Baryon asymmetry is
quantified by the ratio $\eta=\frac{n_b}{n_{\gamma}}$, where $n_b$
is the number of baryons and $n_{\gamma}$ is the number of photons
in the universe. The present value of the asymmetry in the universe
is $\eta \approx 10^{-10}$.  The three conditions postulated by
Sakharov \cite{sakharov} to generate baryon asymmetry are baryon
number violation, CP violation and out of equilibrium evolution of
universe. Many theories have been proposed to explain baryon
asymmetry. Some of them are GUT baryogenesis \cite{dolgov}, electroweak
baryogenesis \cite{dolgov}, leptogenesis \cite{dine} and the Affleck Dine(AD) mechanism \cite{ad}. Most of
these are still inadequate in explaining the value of $\eta$ and a
lot of work remains to be done to get a complete theory of baryon
asymmetry generation.  In
view of the inefficient baryon production in GUTS, Affleck and
Dine \cite{ad} focused on the supersymmetric extension of GUTS to
generate a new mechanism for baryon production based on flat directions. 
In the minimal supersymmetric standard model(MSSM) the number of degrees of freedom are increased 
by virtue of the fact that  bosons and fermions have supersymmetric counterparts. 
The increase in the number of degrees of freedom results  in directions in field space which have virtually no potential. These are known as flat directions and are made up of squarks or sleptons, so they carry baryon or lepton number. During inflation the squarks and sleptons are free to fluctuate along these directions as it costs  little energy and can form condensates with a large baryon or lepton number.
Supersymmetry breaking lifts these flat directions and
sets a scale for the potential.
Supersymmetry breaking can introduce terms that violate baryon number and CP.
In the Affleck Dine model, the cosmological constant in early universe breaks 
the  supersymmetry spontaneously during inflation and 
this gives rise to $B-L$ violation, satisfying Sakharov's first condition. The scalar
fields through the interaction with the inflaton field generate C
and CP violation, thus satisfying  Sakharov's second condition.
  The
expectation values of massless scalar fields can start out displaced
from true minimum, oscillations around the minimum occur when the
Hubble constant becomes comparable to their effective mass resulting in
coherent production of scalar fields manifested as a  condensate of light
scalar particles. This condensate stores baryonic charge and when
inflation is over its decay produces nonzero baryon
asymmetry.

Dine et.al.\cite{dine}showed that baryon asymmetry can be generated for a  scalar field Lagrangian with an interaction term of a general quartic type with  complex couplings
of the form 
\begin{equation*}
L_I=-\lambda|\phi|^4+\epsilon\phi^3\phi^\dagger+\varrho\phi^4+C.C.,
\end{equation*}
$\lambda$, $\epsilon$ and $\varrho$ are of the order of $M_S^2/M_G^2$,
where $M_G$ is the grand unification scale and $M_s$ is the supersymmetry breaking scale.
The baryon number per particle at very large times ($t\gg m_\phi^{-1}$) in both the
radiation and matter dominated eras is given by 
\begin{equation*}
 r \approx \frac{Im (\epsilon+\varrho) \phi_0^2}{m_{\phi}^2},
\end{equation*}
where, $m_\phi$ is of the order of $M_S$ and $\phi_0$ is the vacuum expectation value of the scalar field.
It can be seen that 
if  $\epsilon$ and $\varrho$ are real then the asymmetry vanishes.

This mechanism is too efficient and produces a 
baryon asymmetry that is too large. Attempts to dilute this have
been proposed, but most of these models use classical arguments,
where additional entropy is released after baryogenesis through
decay of the inflaton field \cite{linde, ellis}. Other
models introduce nonrenormalizable terms \cite{ng,din}. In
\cite{yeo}, a preliminary perturbative analysis of out of equilibrium 
quantum fluctuations in the AD model has been shown to lead 
to some amount of reduction in the asymmetry. In this paper
we do a comprehensive study of quantum fluctuations  in a non-perturbative fashion, using the squeezed state formalism, which allows the analysis of quantum effects and expansion on Affleck Dine baryogenesis.

\section{The Formalism}
To carry out our analysis for AD baryogenesis for 
quantum fluctuations arising  post inflation,
we choose a Lagrangian with complex quartic couplings of the form \cite{yeo}
\begin{equation}
S = \int d^4x
\sqrt{-g}[g_{\mu\nu}(\partial^\mu\phi^\dagger)(\partial^\nu\phi)
-m_\phi^2\phi^\dagger\phi - i\lambda(\phi^4-\phi^{\dagger 4})]
\end{equation}
where $\phi$ is a complex scalar field, $m_\phi$ is the mass of the
scalar field and $\lambda$ is a dimensionless real  coupling constant.
The baryon number violation comes from the coupling $\lambda \approx
\epsilon M_S^2/M_G^2$ where $M_S$ is the supersymmetry 
breaking scale, $\epsilon$ is a real parameter which characterises CP
violation and $M_G$ is the grand unification scale. The background metric is considered to be the flat FRW metric.
$$ds^2 = dt^2 -a^2(t)dx^2$$
where a(t) is the expansion parameter.
The classical equation of motion  of the field is,
\begin{eqnarray}
 \ddot{\phi}+3H\dot{\phi}+m^2_\phi\phi^\dagger=4i\lambda\phi^{\dagger 3}\\
\ddot{\phi^\dagger}+3H\dot{\phi^\dagger}+m^2_\phi\phi^\dagger=4i\lambda\phi^{3}
\end{eqnarray}
where H is Hubble's constant.
The initial conditions at $t=t_0$ are given by
\begin{eqnarray}
\phi\vert_{t=t_0}=i\phi_0,~~~\dot{\phi}\vert_{t=t_0}=0.
\end{eqnarray}
The baryon number per particle for large times in this model is 
is given by
\begin{equation}
 r \approx \frac{\lambda \phi_0^2}{m_\phi^2},
\end{equation} in concurrence with the classical AD result.

We now study the quantum effects on $'r'$. We rewrite eqn.[1] in terms of  conformal time $\eta=\int \frac{dt}{(a(t)}$ as
\begin{eqnarray}
S &=& \int d\eta d^3x~
a^4[\frac{1}{a^2}((\frac{\partial\phi^\dagger}{\partial\eta})
(\frac{\partial\phi}{\partial\eta})-(\nabla\phi^\dagger)(\nabla\phi))\nonumber\\
&&-m_\phi^2\phi^\dagger\phi - i\lambda(\phi^4-\phi^{\dagger 4})].
\end{eqnarray}
Defining $\chi=a(\eta)\phi$, we get 
\begin{eqnarray}
S &=& \int d\eta d^3x
[((\frac{\partial\chi^\dagger}{\partial\eta})
(\frac{\partial\chi}{\partial\eta})-(\nabla\chi^\dagger)(\nabla\chi))\nonumber\\
&&-(m_\phi^2a^2-\frac{a''}{a})\chi^\dagger\chi -i\lambda(\chi^4-\chi^{\dagger 4})].
\end{eqnarray}
Decomposing $\chi$ into two real scalar fields,
\begin{eqnarray}
\chi = \frac{1}{\sqrt{2}}(\chi_1 + i\chi_2)\\
\chi^\dagger = \frac{1}{\sqrt{2}}(\chi_1 - i\chi_2),
\end{eqnarray}
and substituting into the action, we get 
\begin{eqnarray}
S&=&\int d\eta d^3x[\frac{1}{2}(\chi'_1)^2 -\frac{1}{2}(\nabla\chi_1)^2
+ \frac{1}{2}(\chi'_1)^2 - \frac{1}{2}(\nabla\chi_2)^2 \nonumber\\&&-
\frac{m_\eta^2}{2}(\chi_1^2 + \chi_2^2) -
2\lambda\chi_1\chi_2[\chi_1^2 -\chi_2^2]],
\end{eqnarray}
where prime denotes derivative with respect to conformal time $\eta$ and $m_\eta^2=m_\phi^2 a^2-\frac{a''}{a}$.

Using  the background field method to study quantum
effects\cite{bm1,bm2,bm3}, we assume field $\chi_i$, $i=1,2$,  has
background classical component $\chi_{i0}$  and a quantum(order $\frac{h}{2\pi}$) field
$\widehat{\chi}_i$:
\begin{equation}
\chi_i=\chi_{i0} +\widehat{\chi}_i
\end{equation}
where $\chi_{i0}$  satisfies the
classical equation of motion,
\begin{equation}
\frac{\delta S}{\delta \chi_i}\vert_{(\chi_i=\chi_{i0})}=0
\end{equation}
and the field $\widehat{\chi}_i$ represents quantum fluctuations around
the classical solution. We expand the action in terms of Taylor series
\begin{eqnarray}
S[\chi_i,\chi_j]&=&S[\chi_i,\chi_j] +\frac{\delta
S[\chi_i,\chi_j]}{\delta \chi_i}\vert_{(\chi_i=\chi_{i0})}\nonumber\\&&
+\frac{1}{2}(
\widehat{\chi}_i\vert
 \frac{\delta^2 S[\chi_i,\chi_j]}{\delta \chi_i\delta \chi_j}
\vert_{(\chi_i=\chi_{i0})}\widehat{\chi}_j)+.........
\end{eqnarray}
Since $\chi_{i0}$ 
satisfies the classical equation of motion, 
the second term is zero and the contribution 
of the  quantum fluctuations comes from the
quadratic and higher order terms. 
\begin{eqnarray}
S&=&\int d\eta d^3x[\frac{1}{2}(\widehat{\chi}'_1)
-\frac{1}{2}(\nabla\widehat{\chi}_1)^2 +
\frac{1}{2}(\widehat{\chi}'_2) -
\frac{1}{2}(\nabla\widehat{\chi}_2)^2 \nonumber\\&& -\frac{
m_\eta^2}{2}(\widehat{\chi}_1^2 + \widehat{\chi}_2^2)-2\lambda(\rho\widehat{\chi}_{1}\widehat{\chi}_{2} + \delta
( \widehat{\chi}_{1}^2-\widehat{\chi}_{2}^2))]
\end{eqnarray}
where $\rho=3(\chi_{10}^2-\chi_{20}^2)$ and $ \delta=3\chi_{10}\chi_{20}$.
 
Using the Legendre transformation, we get the effective Hamiltonian of the fluctuations to be
\begin{eqnarray}
H &=&\int d\eta d^3x [\frac{\widehat{p}_1^2 }{2}+ \frac{(\nabla\widehat{\chi}_1)^2}{2}
+  \frac{m_\eta^2}{2}\widehat{\chi}_1^2+
\frac{\widehat{p}_2^2}{2} + \frac{(\nabla\widehat{\chi}_2)^2}{2}\nonumber\\&&
+ \frac{m_\eta^2}{2}\widehat{\chi}_2^2 +
2\lambda(\rho\widehat{\chi}_{1}\widehat{\chi}_{2} + \delta
( \widehat{\chi}_{1}^2-\widehat{\chi}_{2}^2))]
\end{eqnarray}
 here $\widehat{p}_i$ are the canonical momenta of the $\widehat{\chi}_i$ fields.

Carrying out the mode expansion of the fields, we get 
\begin{eqnarray}
\widehat{\chi}_1 = \int dk[a_k^\dagger e^{ik\cdot x} + a_k e^{-ik\cdot x}],\\
\widehat{\chi}_2 = \int dk[b_k^\dagger e^{ik\cdot x} + b_k e^{-ik\cdot x}],
\end{eqnarray}
where,
\begin{eqnarray}
k\cdot x=k_\mu x^\mu &=&\omega \eta -k_ix_i,\\ \nonumber
d\tilde{k} &=& \frac{d^3kd\eta}{[(2\pi)^3 2\omega]^\frac{1}{2}},\\ \nonumber
\omega^2 &=& k^2 + m_{\eta}^2.
\end{eqnarray}

The mode Hamiltonian is  
\begin{eqnarray}
H &=& \int \frac{d^3k}{(2\pi)^3}[[\frac{\omega}{2}
+\frac{\lambda  \delta}{2\omega}](a_k^\dagger a_k + a_{-k} a_{-k}^\dagger) \nonumber\\&&+
[\frac{\omega}{2}-\frac{\lambda  \delta}{2\omega}](b_k^\dagger b_k + b_{-k} b_{-k}^\dagger) ]\nonumber\\&&+
(\frac{\lambda  \rho}{2\omega}
(a_k^\dagger b_k + a_{-k} b_{-k}^\dagger + a_k^\dagger b^\dagger_{-k}
+a_kb_{-k})\nonumber\\&&
+\frac{\lambda  \delta}{2\omega}
( a_k^\dagger a^\dagger_{-k}
+a_ka_{-k})-\frac{\lambda \delta}{2\omega}
( b_k^\dagger b^\dagger_{-k}
+b_kb_{-k})]].
\end{eqnarray}

To see the symmetries of the Hamiltonian, we define the following generators
\begin{eqnarray} N_{1} = \frac{1}{2}(a_k^\dagger a_k + a_{-k} a_{-k}^\dagger),\; \;
N_{2}=\frac{1}{2}(b_k^\dagger b_k + b_{-k} b_{-k}^\dagger)
 \end{eqnarray}
\begin{eqnarray}
 J_{+} &=& \frac{1}{2}(a_k^\dagger b_k + a_{-k}^\dagger b_{-k}), \;
 J_{-} =\frac{1}{2}(b_k^\dagger a_k  + b_{-k}^\dagger a_{-k}),\nonumber\\
 J_{0} &=&\frac{1}{2}(N_{1} - N_{2}),
\end{eqnarray}

\begin{eqnarray}
 K_{+} = a_k b_{-k},\;
  K_{-} = b^\dagger_{-k} a^\dagger_{k},\;
   K_{0} =\frac{1}{2}(N_{1} + N_{2} + 1),\nonumber\\
  L_{1-} = a^{\dagger}_k a^{\dagger}_{-k},\;
  L_{1+} = a_{-k} a_{k},\;
 L_{10} =\frac{1}{2}(N_{1}  + 1),\nonumber\\
  L_{2-} = b^{\dagger}_k b^{\dagger}_{-k},\;
  L_{2+} = b_{-k} b_{k},\;
 L_{20} =\frac{1}{2}(N_{2}  + 1),.
\end{eqnarray}
 We can show that $(J_{+},J_{-},J_0)$ satisfy an  $su(2)$ algebra
and $(K_{+},K_{-},K_0)$, $(L_{1+},L_{1-},L_{10})$, 
$(L_{2+},L_{2-},L_{20})$ satisfy $su(1,1)$ algebras.

In terms of these generators and the number operators , the Hamiltonian is 
\begin{eqnarray}
H &=& \int \frac{d^3k}{(2\pi)^3}[(\frac{\omega}{2}
+\frac{\lambda  \delta}{2\omega})N_1 +
(\frac{\omega}{2}-\frac{\lambda  \delta}{2\omega})N_2 \nonumber\\&&+
\frac{\lambda  \rho}{2\omega}
(J_+ + J_- + K_+ + K_-)\nonumber\\&&
+\frac{\lambda  \delta}{2\omega}
(L_{1+} + L_{1-})-\frac{\lambda \delta}{2\omega}
( L_{2+} + L_{2-})]
\end{eqnarray}
and in this form explicitly displays $su(1,1)$ and $su(2)$ symmetries.

We diagonalize this Hamiltonian in two steps. First we use a unitary rotation transformation and then a squeezing transformation\cite{Perelomov}. The first transformation is given by
  \begin{equation}
H_1=U^\dagger(R_1) H U(R_1),
\end{equation}
where
\begin{equation} U(R_1)=exp[\theta(J_{+}e^{2i\xi}+
J_{-} e^{2i\xi})],
\end{equation}
The operator $U(R_1)$ provides the following transformation relations:
\begin{eqnarray}
&&U^{\dag}(R_1)\left(\begin{array}{c}a_k
\\b_k\end{array}\right)U(R_1) \nonumber\\&&= \left(\begin{array}{cc}
Cos(\theta)& e^{2i\xi}Sin(\theta)\\-e^{-2i\xi}Sin(\theta)&
Cos(\theta)\\
\end {array}\right)
\left(\begin{array}{c} a_k \\ b_k
\end {array}\right)\nonumber\\&&=
\left(\begin{array} {c} A_k \\ B_k
\end{array}\right),
\end {eqnarray}
the angle $\theta$ is defined from the relation $Sin(2\theta)=(\frac{\rho}{\sqrt{\rho^2+\delta^2}})$ 
The creation and annihilation operators   $A_k$ and $B_k$ are 
\begin{eqnarray}
 A_k= a_k Cos(\theta) + b_k e^{2i\xi}Sin(\theta)\\
B_k= b_k Cos(\theta) - a_k e^{2i\xi}Sin(\theta),
\end{eqnarray} and their complex conjugates.

The Hamiltonian $H_1$ in terms of 
$A_k^\dagger$, $B_k^\dagger$, $A_k$ and $B_k$ is given by
\begin{eqnarray}
 H_1 &=&  \int \omega^2\frac{d^3k}{(2\pi)^3} (
m_1[A_k^\dagger
A_k +A_{-k}^\dagger
A_{-k}]\nonumber\\&&+m_2[B_k^\dagger B_k + B_{-k}^\dagger
B_{-k}]+n_1 [A_k A_{-k} +
A_{-k}^\dagger A_k^\dagger ]\nonumber\\&&+ n_2[ B_k B_{-k} +
B_{-k}^\dagger B_k^\dagger])
\end{eqnarray}
where, $ \Omega^2 =\lambda \sqrt{(\rho^2+\delta^2)}$,
$m_1=\frac{\omega^2 -  \Omega^2}{\omega^2},~~
m_2=\frac{\omega^2 +  \Omega^2}{\omega^2},~~
n_1=  -  \frac{\Omega^2 }{\omega^2}$ and
$n_2=  \frac{\Omega^2}{\omega^2}$. 
 
We again define new generators $(D_{1+},D_{1-},D_{10})$, $(D_{2+},D_{2-},D_{20})$ satisfying $su(1,1)$ algebras.
\begin{eqnarray}
D_{1+}&=&A^\dagger_kA^\dagger_{-k},\; D_{1-}=A_{-k}A_k,\nonumber\\
D_{10}&=&\frac{1}{2}(A^\dagger_k A_k +A^\dagger_{-k} A_{-k} + 1), \nonumber\\
D_{2+}&=&B^\dagger_{-k}B^\dagger_{k},\; D_{2-}=B_{k}B_{-k}\nonumber\\
D_{20}&=&\frac{1}{2}(B^\dagger_{-k} B_{-k} + B^\dagger_{k} B_{k} +1),
\end{eqnarray}
and rewrite the  Hamiltonian in terms of the  new generators
\begin{eqnarray}
H_1&=& \int \frac{d^3k}{(2\pi)^3}\omega^2[[m_1
D_{10}  +m_2 D_{20}]
\nonumber\\&&+n_1[D_{1+}
+D_{2-}]+n_2
(D_{2+}+D_{1-})],
\end{eqnarray}
showing su(1,1) symmetry. We can diagonalize the Hamiltonian using
squeezing (Bogolubov) transformation
\begin{equation}
H_f=S(\zeta_2)^\dagger S(\zeta_1)^\dagger H_1 S(\zeta_1)S(\zeta_2)
\end{equation}
where $S(\zeta_1)=exp[\zeta_1D_{1+}
-\zeta_1^* D_{1-}],~S(\zeta_2)=exp[\zeta_2 D_{2+} - \zeta^*_2D_{2-}]$, $\zeta_1=r_1exp[i\gamma]$ and $\zeta_2=r_2exp[i\gamma]$. 

The effect of the  operators $S(\zeta_1)$ and $S(\zeta_2)$ on  $A_k$ and $B_k$ is 
\begin{eqnarray}
A_s(k,\eta) = \mu_1 A_k + \nu_1 A^\dagger_{-k},\\
A_s^\dagger(k,\eta) = \mu_1^* A_k^\dagger + \nu_1^* A_{-k},\\
B_s(k,\eta) = \mu_2 B_{-k} + \nu_2 B^\dagger_{k},\\
B_s^\dagger(k,\eta) = \mu_2^* B_{-k}^\dagger + \nu_2^* B_{k},
\end{eqnarray}
where $\mu_1=Cosh(r_1)=\frac{m_1}{\sqrt{m_1^2-n_1^2}}$,
$\nu_1=e^{-i\gamma}Sinh(r_1)=e^{-i\gamma}\frac{n_1}{\sqrt{m_1^2-n_1^2}}$,
$\mu_2=Cosh(r_2)=\frac{m_2}{\sqrt{m_2^2-n_2^2}}$ and
$\nu_2=e^{-i\gamma}Sinh(r_2)=e^{-i\gamma}\frac{n_2}{\sqrt{m_2^2-n_2^2}}$.
Thus the final diagonalized Hamiltonian  after two unitary transformations is 
\begin{eqnarray}
H_f&=& \int \frac{d^3k}{(2\pi)^3}
\Omega_{+}[A^\dagger_s(k,\eta)A_s(k,\eta)+1]\nonumber\\
&& +\Omega_{-}[B^\dagger_s(k,\eta)B_s(k,\eta)+1]],\label{h}
\end{eqnarray}
where $\Omega_+ = \sqrt{m_1^2-n_1^2}=\sqrt{\omega^2 -  2\Omega^2}$ 
and $\Omega_ = \sqrt{m_2^2-n_2^2}=\sqrt{\omega^2 +  2\Omega^2}$

The vacuum state of $H_f$ at time $\eta$ is given by $\vert 0(\eta),0(\eta)>$ 
and  vacuum state of $H$ at initial time is given by $\vert 0,0>$, which are related by
\begin{equation}
 \vert 0(t),0(t)>=e^{\int \frac{d^3k}{(2\pi)^3}[\zeta_1D_{1+}
-\zeta_1^* D_{2-}][\zeta_1D_{1+}
-\zeta_1^* D_{2-}]} \vert 0,0>
\end{equation} 
We see that the vacuum at later times  is populated with particles and anti-particles
with respect to vacuum state at initial time.
We can estimate the number of particles and
anti-particles from the relationship between the creation and
annihilation operators at initial time given by $a_k$,
 and $b_k$, and the final creation and annihilation at later time given by
operators $A_s$ and $B_s$.
\begin{eqnarray}
A_s(k,\eta) &=& (\mu_1Cos(\theta))a_k + (\nu_1Sin\theta e^{2i\xi}) b_{-k}^{\dagger}
\nonumber\\&&+ (\mu_1 Sin\theta e^{2i\xi}) b_k + (\nu_1Cos(\theta)) a_{-k}^{\dagger},\\
B_s(k,\eta) &=& (\mu_2Cos(\theta)) a_{-k} + (\nu_2Sin\theta e^{-2i\xi})  b_{k}^{\dagger}
\nonumber\\&&+ (\mu_2 Sin\theta e^{-2i\xi})  b_{-k} + (\nu_2Cos(\theta)) a_{k}^{\dagger}.
\end{eqnarray}
The number of particles (baryons) for each mode
\begin{equation}
N_{kB}(\eta) = \langle B_s^\dagger(k,\eta) B_s(k,\eta)\rangle = \vert\nu_{k2}\vert^2,
\end{equation}
and the number of anti-particles (anti-baryons) for each mode
\begin{equation}
 N_{k\overline{B}}(\eta)= \langle A_s^\dagger(k,\eta) A_s(k,\eta)\rangle  = \vert\nu_{k1}\vert^2.
\end{equation}

Therefore the baryon asymmetry is
\begin{eqnarray}
\triangle N_k=N_B(\eta)-N_{\overline{B}}(\eta) 
&=&\frac{\Omega^6}{\omega^2(4\Omega^4-\omega^4)}
\end{eqnarray}
where $ \Omega^2 =\lambda \sqrt{(\rho^2+\delta^2)}$, recalling that $\rho=3(\chi_{10}^2-\chi_{20}^2)$ and $ \delta=3\chi_{10}\chi_{20}$. We find that  $N_B(\eta)-N_{\overline{B}}(\eta)$ is dependent on vacuum expectation values of real and imaginary parts of scalar field and coupling constant $\lambda$.

The total asymmetry is given by,
\begin{eqnarray}
\int_0^\infty\triangle N_k d^3k
&=&\int_0^\infty k^2 dk\frac{\Omega^6}{\omega^2(4\Omega^4-\omega^4)}\nonumber\\
&=&|\Omega^2 (\sqrt{m_\eta^2+2\Omega^2}-\sqrt{m_\eta^2-2\Omega^2})|
\end{eqnarray}
It is interesting to see that when $\lambda\ll 1$ and 
$\chi_{10}=\chi_{20}={\phi_0}$  
the asymmetry reduces to classical value,

\begin{eqnarray}
(N_B(\eta)-N_{\overline{B}}(\eta)) &=&(\frac{3\lambda
\phi_0^2}{4m_\eta^2})\simeq r.
\end{eqnarray}

\section{Evolution Of Asymmetry Parameter:}
In order to get some exact results and numerical values for the  parameter $r$ after expansion,
 we consider a (quite realistic) expansion where we can evaluate the Bogolubov
 coefficients exactly.

For this consider the time evolution of wave function
under the action of the Hamiltonian $H_f$ given by (\ref{h}).
Going over to the coordinate representation $\Pi_{(A,B)}$ and $P_{(\Pi_{A,B})}$ defined by the relations \cite{bbkvs},
$ A_s(k,\eta)=\frac{e^{i\int\Omega_+(\eta)d\eta}}{2\sqrt{\Omega_+(\eta)}}
(\Omega_+(\eta)\Pi_A(k,\eta) + iP_{\Pi_A}(k,\eta))$ and $
B_s (k,\eta)=\frac{e^{i\int\Omega_-(\eta)d\eta}}{2\sqrt{\Omega_-(\eta)}}
(\Omega_-(\eta)\Pi_B (k,\eta)+iP_{\Pi_B}(k,\eta) )$ ( and their complex conjugates),
the Hamiltonian $H_{f}$ is
\begin{equation}
H_f(\eta)=\int\frac{d^3k}{(2\pi)^3}\sum_{i=A,B}\frac{1}{2}[(\Omega_+)^2\Pi^2_i(k,\eta)+P_{\Pi_i}^2(k,\eta)]
\end{equation}

The time evolution of a wave function $\psi(\eta)$ under the action of a Hamiltonian $H(\eta)$ is simply
\begin{equation}
H(\eta)\psi(\eta)=i\frac{d}{d\eta}\psi(\eta),
\end{equation}
From the form of $H_f$ given above, it is clear that it is the
direct sum of two independent Hamiltonian $ H_{A}(\eta)$ and $ H_{B}(\eta)$
for each of the $A_s$ and $B_s$ modes. The wave function
 for each momentum mode evolves as
 \begin{eqnarray}
H_{A}(k,\eta)\psi_{A}(k,\eta)=i\frac{d}{d\eta}\psi_{A}(k,\eta),\\
H_{B}(k,\eta)\psi_{B}(k,\eta)=i\frac{d}{d\eta}\psi_{B}(k,\eta).
\end{eqnarray}
Since the Hamiltonians $H_{A}$ and $H_{B}$ are time dependent
harmonic oscillators, in the coordinate space representation 
the wave functions $\psi_{A}(k,\eta)$
and $\psi_{B}(k,\eta)$ can be represented by gaussian wave functions.

After some algebra, the evolution equations satisfied by the two wave functions for each mode are
\begin{eqnarray}
\psi''_{A}(k,\eta)+\Omega^2_-\psi_{A}(k,\eta) &=& 0,\\
\psi''_{B}(k,\eta) +\Omega^2_+\psi_{B}(k,\eta) &=& 0.
\end{eqnarray}
where $\Omega_- =  k^2  + m_\phi^2a^2 -\frac{a''}{a}-2\lambda \sqrt{\rho^2+\delta^2}$ 
and $\Omega_+ = k^2  + m_\phi^2a^2 -\frac{a''}{a}+2\lambda \sqrt{\rho^2+\delta^2}$

We rewrite these equations in the {\it Schroedinger like form} in $\eta$

\begin{equation}
\psi_{A}'' + ( E + V_1(a) )\psi_{A} =0,\label{df1}
\end{equation}
and
\begin{equation}
\psi_{B}'' +( E + V_2(a))\psi_{B} =0,\label{df2}
\end{equation}
where 
where 
\begin{eqnarray}
E = k^2  + m_\eta^2
\end{eqnarray}
\begin{eqnarray}
V_1(a) &=& -2\lambda \sqrt{\rho^2+\delta^2},\\
V_2(a) &=& 2\lambda \sqrt{\rho^2+\delta^2}.
\end{eqnarray}
where $\sqrt{\rho^2+\delta^2}=3\sqrt{\chi_{10}^4+\chi_{20}^4-\chi_{10}^2\chi_{20}^2}$.
Writing the equations in this form allows us to use the machinery of potential barrier reflection and transmission problems in quantum mechanics.
The reflection ($R$) and transmission ($T$) coefficients can be related to the
squeezing parameter ($r$) through the relation $sinh^2(r)=|\nu|^2=\frac{R}{T}$ allowing us to calculate  N(k) =$|\nu|^2$. We can also explicitly see the origin of the  asymmetry in the amplification of the particle and anti-particle modes. The particles face a potential barrier and the antiparticles a potential well. This results in a differential evolution of the particle and antiparticle modes resulting in baryon asymmetry.

We now see that there are two factors that contribute to the time evolution of the particle and anti-particle modes, the time dependence of the background classical solution and the time dependence of the expansion factor '$a(\eta)$'.
\subsection{Slow Expansion}
As a first approximation we consider the case when $\frac{a''}{a}=0$, i.e , radiation dominated universe. In this case as seen in \cite{ad} the classical equations for $\chi_{10}$ and $\chi_{20}$ are given by
\begin{equation}
\chi_{10}'' + m_\eta^2\chi_{10} =6\lambda \chi_{10}^2\chi_{20}-2\lambda\chi_{20}^3,
\end{equation}
and
\begin{equation}
\chi_{20}'' + m_\eta^2\chi_{20} =-6\lambda \chi_{20}^2\chi_{10}+2\lambda\chi_{10}^3
\end{equation}
To solve these equations analytically we neglect the 
cubic term and then we apply the boundary conditions 
in equation (4),to get the time dependent background solutions
 \begin{eqnarray}
\chi_{10}&=&\frac{\lambda\phi_0}{(m_\eta)^3}sin(m_\eta\eta+\epsilon)\label{c}\\
\chi_{20}&=&\frac{\phi_0}{m_\eta}sin(m_\eta\eta)\label{c1}.
\end{eqnarray}

Then to upto first order  we have  $\lambda \sqrt{\rho^2+\delta^2}=\overline{\varphi}^2sin^2(m_{\eta}\eta)$ 
 where $\overline{\varphi}$ is a slowly decreasing amplitude,  given by 
$\overline{\varphi}^2=3\lambda(\frac{\phi_0}{m_\eta})^2$. 

The equations \ref{df1} and \ref{df2} become
\begin{equation}
\psi_{A}'' + ( k^2 +m_\eta^2 - \overline{\varphi}^2
sin^2(m_{\eta}\eta))\psi_{A} =0,\label{s}
\end{equation}
and
\begin{equation}
\psi_{B}'' +( k^2 +m_\eta^2 + \overline{\varphi}^2
sin^2(m_{\eta}\eta))\psi_{B} =0.\label{s1}
\end{equation}.

These can be written as the Mathieu equations associated which are  familiar from
the parametric amplification problem in inflation.
\begin{equation}
\psi_{A}'' + \omega_{1k}^2( 1 + \frac{\overline{\varphi}^2}{\omega_{1k}^2}
cos(\gamma\eta)-\frac{a''}{\omega_{1k}^2a})\psi_{A} =0,\label{su1}
\end{equation}
and
\begin{equation}
\psi_{B}'' +\omega_{2k}^2( 1 - \frac{\overline{\varphi}^2}{\omega_{2k}^2}
cos(\gamma\eta)+\frac{a''}{\omega_{2k}^2a})\psi_{B} =0.\label{su2}
\end{equation}.
where $ \omega_{1k}^2=k^2+m^2_\phi-\overline{\varphi}^2$, 
$ \omega_{2k}^2=k^2+m^2_\phi+\overline{\varphi}^2$ and $\gamma=2m_\eta$

To solve these equations we follow the method given in \cite{landau} and \cite{kaiser}.
From the theory of parametric resonance, the resonance is strongest if the frequency 
is twice $\omega_{ik}$, hence we put $\gamma=2\omega_{ik}+\varepsilon$ with $\varepsilon \ll \omega_{ik}$, 
 the resonance condition will be satisfied if $
\overline{\varphi}^2 - \varepsilon^2>0$
or $
\vert\varepsilon\vert<\overline{\varphi}$.
We define a new variable $
l=\frac{\varepsilon}{\overline{\varphi}}$
so that resonance occurs $-1<l<1$.

Then by using \cite{kaiser}, the  number of particles produced is given by
\begin{equation}
N_{1k}=\frac{1}{1-l^2}Sinh^2(\sqrt{\overline{\varphi}^2-\varepsilon^2}\eta),
\end{equation}
and the number of anti-particles is 
\begin{eqnarray}
N_{2k}=\frac{1}{1-l^2}Sin^2(\sqrt{\overline{\varphi}^2-\varepsilon^2}\eta)
\end{eqnarray}
\begin{figure*}
 \subfigure[($N_1(k)-N_2(k)$) and $\overline{\varphi}= 10^{-2}e^{-\eta/3}$]{\epsfig{file=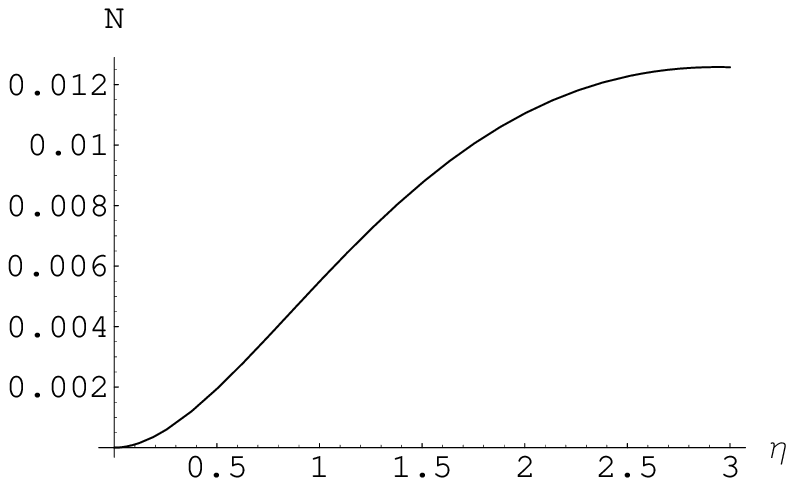,scale=0.5}}\vspace{.5in}
\subfigure[($N_1(k)-N_2(k)$) and  $\overline{\varphi}= 10^{-3}e^{-\eta/5}$]{\epsfig{file=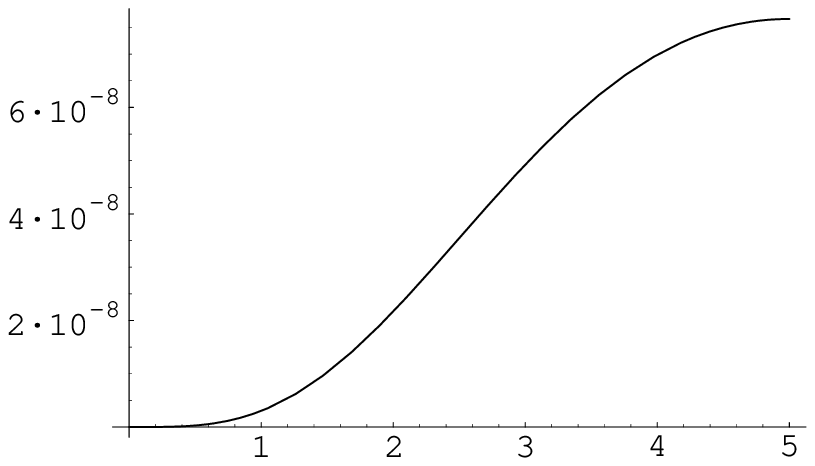,scale=0.5}}
 \end{figure*}

For parametric resonance, it is important that the inflaton stays in resonance band and this is possible as long as its amplitude is slowly varying function of time. The time dependence of the number of particles and antiparticles comes from the slow time variation of the decaying amplitude, which we phenomenologically  approximate with 
$\overline{\varphi}^2\simeq \overline{\varphi}^2e^{-\eta/\tau}$ where $\tau$ is damping scale.
In the figure (a) and (b) we have plotted $N_1(k)-N_2(k)$ for various values of the values of $\overline{\varphi}$.
We can see that
the value of asymmetry saturates to a finite value of $\approx=10^{-8}$.

We assume broad resonance such that the Mathieu equation has 
instability bands with in which parametric resonance occurs,
we shall select the first instability region as broad resonance band.

Therefore in the region of broad band resonance we replace the oscillating potential near its zeros
with an asymptotically flat potential of the form
\begin{eqnarray}
 |\overline{\varphi}|^2sin^2(m_{\phi}\eta)\simeq
2|\overline{\varphi}|^2tanh^2(m_{\phi}\frac{(\eta-\eta_i)}{\sqrt{2}}).
\end{eqnarray}
Then \ref{df1}and \ref{df2} become
\begin{equation}
\psi_{A}'' + ( k^2 +m_\phi^2 - \overline{\varphi}^2
tanh^2(\frac{m_{\phi}(\eta-\eta_i)}{\sqrt{2}}))\psi_{A} =0,
\end{equation}
and
\begin{equation}
\psi_{B}'' +( k^2 +m_\phi^2 +\overline{\varphi}^2
tanh^2(\frac{m_{\phi}(\eta-\eta_i)}{\sqrt{2}}))\psi_{B} =0.
\end{equation}.

We get the following differential equations for the particle and antiparticle modes
\begin{equation}
\frac{d^2\psi_A}{dy^2} + \left[\kappa_{1}^2 +
\overline{\varphi}^2 sech^2(y)\right]\psi_A=0.
\end{equation}\begin{equation}
\frac{d^2\psi_B}{dy^2} + \left[\kappa_2^2 +
\overline{\varphi}^2tanh^2(y)\right]\psi_B=0.
\end{equation}
where
\begin{eqnarray}
\kappa_{1}^2 &=&\frac{k^2-\overline{\varphi}^2}{m_{\phi}^2}+1,  \nonumber\\
 \kappa_2^2 &=& \frac{k^2}{m_{\eta}^2}+1\\ \nonumber
    y&=&m_{\phi}(\eta-\eta_i).
\end{eqnarray}

Using the transmission and reflection coefficients \cite{Flugge},
the number of particles is  
\begin{eqnarray}
n_{1k} = \vert\nu_{1k}\vert^2
=\frac{\left(cos^2(\pi\sqrt{\overline{\varphi}^2 +
\frac{1}{4}})\right)^2}
{\left(sinh^2(\pi\kappa_1\right))^2},
\end{eqnarray}
the number of antiparticles is 
\begin{eqnarray}
n_{2k} = \vert\nu_{2k}\vert^2 =
\frac{\left(cosh(\pi\sqrt{\overline{\varphi}^2 -
\frac{1}{4}})\right)^2}
{sinh^2\left(\pi\sqrt{(\overline{\varphi}^2+\kappa_2^2)}\right)
}.
\end{eqnarray}

In the figure (c)and (d)  the evolution of particles and antiparticles for different values of $\overline{\varphi}=.27$ 
$\overline{\varphi}=.28$, $\overline{\varphi}=.29$ are plotted respectively.
We can see clearly that the number of particles increases and 
number of antiparicles decreases due to differential amplification
of particle and antiparicle modes.

\begin{figure*}
 \subfigure[$N_1(k)$ particles(solid line) and $N_2(k)$ and antiparticles(dash line)$\overline{\varphi}=.27$ ]{\epsfig{file=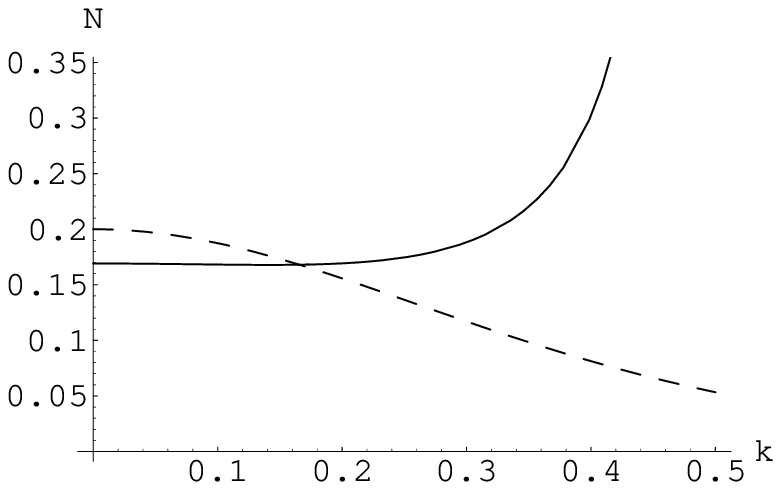,scale=0.5}}\vspace{.5in}
 \subfigure[$N_1(k)$ particles(solid line) and $N_2(k)$ and antiparticles(dash line)$\overline{\varphi}=.28$]{\epsfig{file=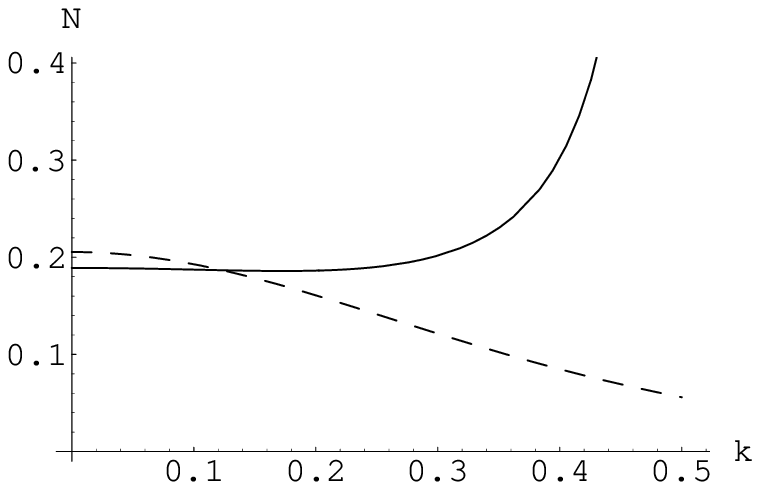,scale=0.5}}\vspace{.5in}
  \end{figure*}
\subsection{Rapid Expansion}
Now we consider the effect of rapid expansion on the asymmetry parameter. We consider the case when the rate of expansion dominates over the oscillation period of the classical background solution so that $\rho $ and $\delta$ can be considered as time independent.
We consider 
\begin{eqnarray}
a(\eta) = (a_0\eta)^\frac{p}{1-p}~~~~~~\eta < \eta_0\\
a(\eta) = C(\eta-\eta_0) ~~~~~~\eta > \eta_0
\end{eqnarray}
where $\eta_0=\eta_*-(a_0^2\eta_*)^{-1}$. It is convenient to set $a(\eta_0)=1$
which sets $\eta_*=a_0^{-1}$, and thus $\eta_0=0$  and $C=a_0^{-1}$ where $a_0=-H_0$
for de Sitter spacetime. Here p=$\frac{1}{2}$ corresponds to radiation dominated universe
and $p\rightarrow \infty$ corresponds to de Sitter epoch.

Thus \ref{df1} is
\begin{eqnarray}
\frac{d^2\psi_A}{d\eta^2}  + \left[k^2 - 2\lambda \sqrt{\rho^2+\delta^2}-\frac{p(2p-1)}{(p-1)^2\eta^2}+\frac{m^2}{H^2\eta^2}\right]\psi_A=0\nonumber\\
\eta < \eta_0 \label{el}
\end{eqnarray}
\begin{equation}
\frac{d^2\psi_A}{d\eta^2}  + g_1\psi_A=0\;\;\eta > \eta_0  \label{dm}
\end{equation}

The equation \ref{el} can be written as
\begin{equation}
\frac{d^2\psi_A}{d\tau_1^2}  + \left[\frac{\frac{1}{4}-q^2}{\tau_1^2}+1\right]\psi_A=0
\end{equation}
where $ q^2=\frac{(3p-1)^2}{4(p-1)^2}-\frac{m_\phi^2}{H}$, $\tau_1=\sqrt{g_1}\eta$
and $ g_1=k^2 - 2\lambda \sqrt{\rho^2+\delta^2}$.
the solution is given by
\begin{equation}
\psi_A = (\sqrt{g_1}\eta)^\frac{1}{2}[A_k H_q^{(1)}(\sqrt{g_1}\eta) + B_k H_q^{(2)}(\sqrt{g_1}\eta)]
\end{equation}
where $H_q^{(1)}$ and $H_q^{(2)}$ are Hankel functions.

The solution for \ref{dm} is given by
\begin{eqnarray}
\psi_A(\eta>\eta_0)=\frac{1}{\sqrt{2k}}[\mu e^{-i\sqrt{g_1}\eta} + \nu e^{i\sqrt{g_1}\eta}]
\end{eqnarray}

The Bogolubov coefficients are obtained by matching the wave functions
and its first derivative at $\eta=\eta_0$

The number of particles is given by.
\begin{eqnarray}
N_1(k)=\vert \nu_1\vert^2= 4^{q-2}(\frac{\sqrt{g_1}}{a_0})^{-2q-1}(q-\frac{1}{2})^2\Gamma^2(q).
\end{eqnarray}

Using similar methods for \ref{df2} the number of  anti particles produced is given by
\begin{eqnarray}
N_2(k)=\vert \nu_2\vert^2=4^{q-2}(\frac{\sqrt{g_2}}{a_0})^{-2q-1}(q-\frac{1}{2})^2\Gamma^2(q).
  \end{eqnarray}
where $ g_2=k^2 + 2\lambda \sqrt{\rho^2+\delta^2}$,
$a_0$ is the reference scale of $H_0$, $H_0$ is constant for de Sitter expansion.

\begin{figure*}
 \subfigure[($N_1(k)-N_2(k)$) $q =.6$ (solid line),$q=.7$ (dash line)  here $\eta=1$ 
$\lambda=10^{-2}$ and $\phi_0=10^{-2}$ ]
 {\epsfig{file=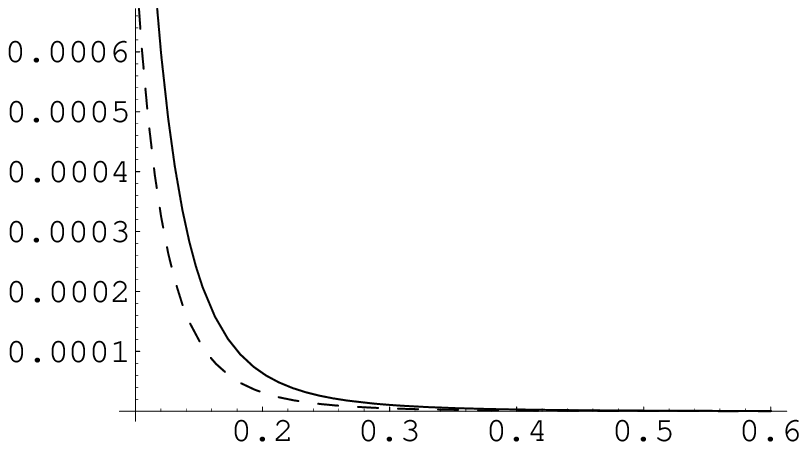,scale=0.5}}\hspace{0.35in}
 \hspace{0.35in}
\subfigure[($N_1(k)-N_2(k)$) here $q =.6$, $\eta=1$ 
$\lambda=10^{-3}$    $\phi_0=10^{-3}$(dash line) and $\phi_0=10^{-4}$ (solid line)]{\epsfig{file=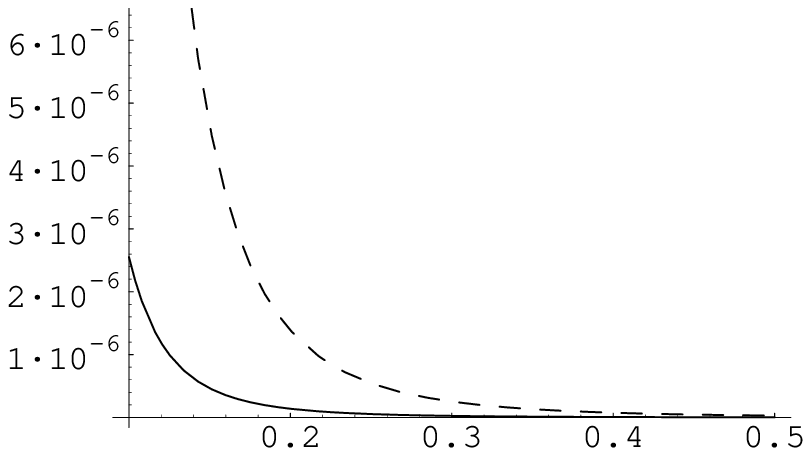,scale=0.5}}
\hspace{0.35in}
\subfigure[$q =.4$, $\eta=1$, $\overline{\varphi}=10^{-2}$ 
and $\lambda=10^{-6}$(dash line),$\lambda=10^{-7}$(solid line)]{\epsfig{file=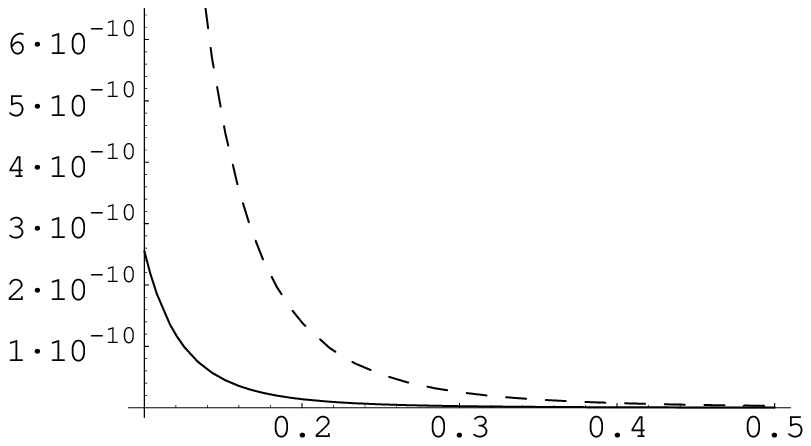,scale=0.5}}
\hspace{0.35in}
\subfigure[($N_1(k)-N_2(k)$) here $q =.4$, $\eta=1$ 
$\lambda=10^{-3}$    $\phi_0=10^{-3}$ (solid line) and $\phi_0=10^{-4}$ (dash line)]{\epsfig{file=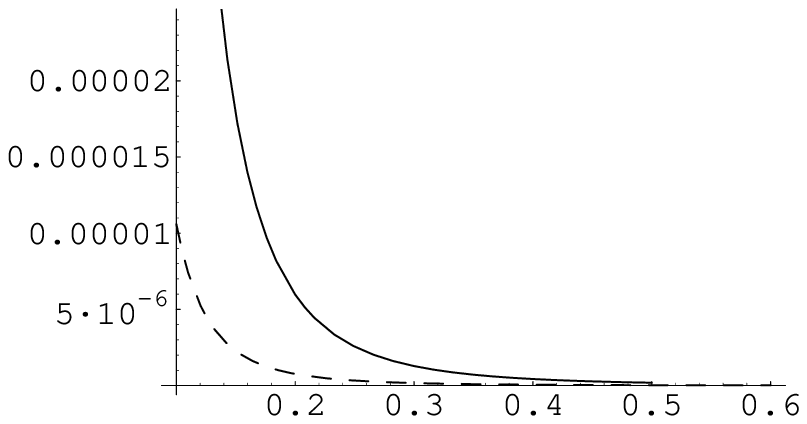,scale=0.5}}
\hspace{0.35in}
\subfigure[$q =.4$, $\eta=1$, $\overline{\varphi}=10^{-2}$ 
and $\lambda=10^{-6}$(dash line),$\lambda=10^{-7}$(solid line)]{\epsfig{file=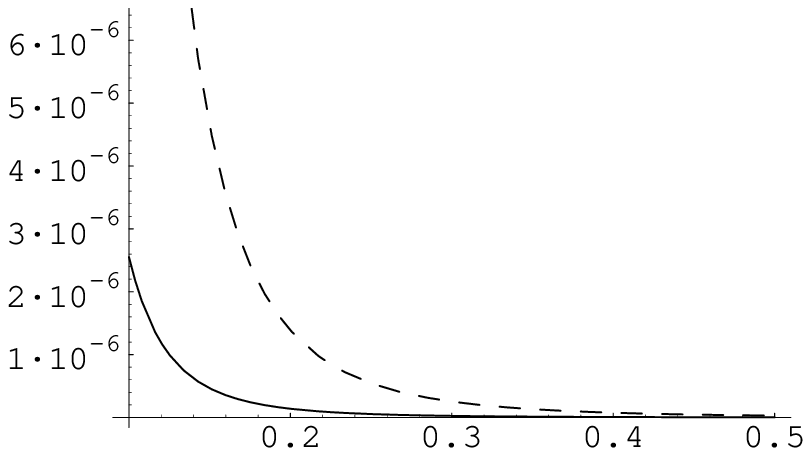,scale=0.5}}
 \end{figure*}

First we have considered a case when $p\rightarrow\infty$ which corresponds to de Sitter epoch.

The vacuum fluctuations of massive fields on exact de Sitter
background leads to density perturbations only for 
$\frac{m_{\phi_0}^2}{H}< 2$ and $2<m^2<\frac{9H_0^2}{4}$.
The corresponding characteristic  values are $q=0$,
and $\frac{m^2}{H^2}=\frac{9}{4}$ for de Sitter case. 
In the figure (e) ($N_1(k)-N_2(k)$) is plotted, with $q=.6$ , $q=.7$,
$q=.8$, which corresponds
to $\frac{m_{\phi_0}^2}{H}<2$ for de Sitter epoch at this value
the condensate starts oscillating and gives rise to fluctuations.

In the figure () $N_1(k)-N_2(k)$ is plotted for the different values of $\phi_0$ for a fixed $q=.6$.
From the figure we can see that as the $\phi_0$ value decreases 
the value of asymmetry reduces.
In the figure (g) $N_1(k)-N_2(k)$ is plotted for the different values of $\lambda$ for a fixed $q=.6$ and  $\phi_0$.
From the figure we can see that as the $\lambda$ value decreases 
the value of asymmetry reduces.
 Large occupation number in a given mode means that quasi-particles formed a condensate.
Therefore from the figure we can see that once the quantum fluctuations are switched on
the value of asymmetry reduces but does not goes to zero.

Now we consider the case when $p\rightarrow\frac{1}{2}$ which corresponds to radiation dominated universe.

In the radiation dominated universe, when 
$\frac{m_{\phi_0}^2}{H}< .25$, then only the vacuum fluctuations 
of massive fields will be switched on, and the 
 characteristic  values for $q=0$,
is $\frac{m^2}{H^2}=\frac{1}{4}$ .

In the figure (h) $N_1(k)-N_2(k)$ is plotted for the different values of $\phi_0$ for a fixed $q=.4$.
corresponds to  $\frac{m_{\phi_0}^2}{H}=.16$.
From the figure we can see that once the quantum fluctuations are switched on
the value of asymmetry reduces.
In the figure (i) $N_1(k)-N_2(k)$ is plotted for the different values of $\phi_0$ for a fixed $q=.4$
for quantum fluctuations, in this case the asymmetry goes to $10^{-3}\frac{\lambda\phi_0^2}{m^2_\phi}$.
We conclude that the Affleck Dine mechaniam when combined with the CP violating amplifacation of vacuum flucutations during inflation can give an acceptable value of baryon asymmetry of universe. 

\section{Conclusion}

In this paper we have studied the non-equilibrium quantum effects of 
Affleck Dine baryogenesis in the post inflationary scenario.
In the paper \cite{yeo} they have studied the model
using the nonequilibrium dynamics and used  perturbative methods to 
get the asymmetry, whereas the methodology developed 
here using squeezed states or
Bogolubov transformations has allowed us to derive the general evolution equations for
baryon and anti-baryon modes 
non-perturbatively in an expanding FRW metric. 
In our evolution equation the effective potential is
dependent upon the expansion parameter $a(\eta)$ and inflaton potential.  
The amount of particle
production in the de Sitter expansion is calculated as the tunnelling
through a barrier of potential $V(a)$. We find that by considering different inflationary scenarios and parametric resonance we can control the asymmetry parameter "r" to a much lower value than in the classical Affleck Dine model. Of course, we have used a simplified toy model, but the method is general and can work for a more realistic scenario also.

Baryon asymmetry remains an  intriguiging unsolved issue in physics. In view of the standard model being in sufficient to explain this essential fact about the universe, one has to look beyond the standard model. Supersymmetry is a compelling idea beyond standard model. Hence finding arguments for baryon asymmetry in MSSM is a natural idea. The Affleck Dine mechanism which uses the flat directions in supersymmetric models is therefore a very promising mechanism for baryon asymmetry generation. Furthermore there is also very compelling evidence that universe went through an inflationary phase and particles were generated by reheating processes. Thus combining the two we should be able to get a plausible and viable scenario for baryon asymmetry. Means to reduce the rather over efficient generation of baryon asymmetry in Affleck Dine mechanism require a th0rough study. In this paper we have done a systematic study of the effects of inflation and parametric amplification of quantum fluctuations on the baryon asymmetry generated in post-inflationary Affleck Dine baryogenesis. Since a variant of the Affleck Dine mechanism is also used to account for dark matter as well as in most exotic scenarios of baryogenesis our method should be useful in this context also.
\begin{acknowledgments}
This work was supported by SAP/CAS, U.G.C and 
as well as Revised fellowship scheme, U.G.C.
 KVSSC and BAB  thank Prof C. Mukku for his valuable comments on this paper.
\end{acknowledgments}

 %Just because of unusual number of tables stacked at end
\bibliography{adk}% Produces the bibliography via BibTeX.

\end{document}